\documentclass{aa}  

\usepackage{graphicx}
\usepackage{txfonts}
\usepackage{natbib,twoopt}
\usepackage{graphicx, xcolor}
\usepackage{float}
\usepackage{txfonts}
\usepackage{multirow}
\usepackage{tablefootnote}
\usepackage{comment}
\usepackage[toc]{appendix} 
\usepackage{array, bm, amsmath}

\newcommand{\BP}{$\beta$~Pic~}

\begin{document}

   \title{New exocomets of \BP}

   \author{Pavlenko Ya.
          \inst{1,2}
          \and
          Kulyk I.\inst{1}
           \and
           Shubina O.\inst{1,3}
          \and
          Vasylenko M.\inst{1} 
          \and
          Dobrycheva D.\inst{1}
          \and
          Korsun P.\inst{1}
          }

   \institute{Main Astronomical Observatory of the NAS of Ukraine, 27, Akademik Zabolotny Str., Kyiv, 03143, Ukraine\\
         \and
             Centre for Astrophysics Research, University of Hertfordshire, College Lane, Hatfield, AL10 9AB, United Kingdom\\
         \and
             Astronomical Observatory of Taras Shevchenko National University of Kyiv, 3 Observatorna Str., 04053 Kyiv, Ukraine\\
             }

   \date{Received --; accepted --}

  \abstract
  {}
   {The aim of our work is to analyze the light curves of \BP recently observed by TESS in sectors 32, 33, and 34, searching for the signatures of exocomet transits. }
   {We process the \BP light curves from the MAST database, applying the frequency analysis to remove harmonic signals due to the star’s pulsations, and use a simple 1-D model to fit the profiles of the found events.}
   {We recover events previously found by other authors in sectors 5 and 6 and find five new distinct aperiodic dipping events with asymmetric shapes resembling the expected profiles due to the passage of a comet-like body across the star disk. These dips are rather shallow, with the flux drop at a level of 0.03\% and a duration less than 1 day. No periodic transits were found in the sectors investigated.}
   {The depth and duration of the identified dips are similar to the recently discovered transits in the \BP light curves from  sector 5 of the TESS observations as well as to those found in the light curves of KIC 354116 and KIC 1108472 from the $Kepler$ database. It indicates that aperiodic shallow dips are not likely an exceptional phenomenon, at least for the \BP system.}


   \maketitle
%

\section{Introduction}

\BP belongs to the \BP moving group, a kinematic group of stars in the vicinity of the Solar system \citep{Zuck2004}. The age of the group population is estimated in the range from about 10 Myr to 40 Myr with a medium consensus age of $23 \pm 3$\,Myr \citep{mama14}. \BP is a young A5V star with a mass of about 1.7\,-- 1.8\,$M_{\odot}$, which is already evolved to be at Zero Age Main Sequence (ZAMS) or on the ZAMS \citep{grif97}. The significant excess in the infrared emissions observed by IRAS (Infra Red Astronomical Satellite) reveals the existence of dust shells around \BP \citep{Backman1986,Cote1987}. 
Successive models of the IR emitting region around \BP suggest its complex architecture, which is confirmed by direct imaging of the ring features and the discovery of two planets \BP b and \BP c, embedded in the disk \citep{Terr1984,Kalas2000,Wahh2003,Lagr2020}. The indirect evidence for the existence of a minor body population, planetesimals and/or cometesimals, in the \BP system was obtained by discovering ``Falling Evaporating Bodies'', FEBs, which manifest themselves as the strong red-shifted variations in the profiles of the circumstellar absorption lines. These variable spectral features can be associated with comet-like bodies in the star-grazing orbits falling on the parent star \citep{ferlet1987,Beust1996}. The red-shifted spectral components can be divided into two distinct classes: low velocity and high velocity features. The former ones are usually deep, and their red-shift velocities with respect to the central component fall between 10-20\,km\,s$^{-1}$and 50\,km\,s$^{-1}$. The latter ones appear to be shallower, varying on a shorter time scale. Their typical red-shifts are about 100\,km\,s$^{-1}$ \citep{Beaust1998}. A strong transient blue-shifted feature was also found in the circumstellar \ion{Ca}{II} K line shifted by 14\,km\,s$^{-1}$ in respect to the `stable', circumstellar component at a velocity of 22\,km\,s$^{-1}$ \citep{Lagr1988,Craw1998}. The blue-shifted feature implies the existence of a comet-like object on a significantly different orbit from those causing the red-shifted variations\citep{Kiefer2014}. The short-term optical variation of the \BP brightness was firstly reported by \cite{Lecavelier1995}. To explain the phenomenon, the mechanism of the star's light scattering by an elongated dust cloud or by a cometary coma passing across the star's disk was adopted as the most plausible one \citep{Lamers1997}. 
Recently \cite{zieba2019} presented the \BP light curve analysis based on the Transiting Exoplanet Survey Satellite (TESS) data. They found three transit events interpreted as exocomet passages across the stellar disc of \BP.

The purpose of this paper is to carry out further investigations of the light curves of \BP, searching for transit events that can be interpreted as exocomet passages across the star's disk based on data from sectors 32, 33, and 34 observed by the TESS recently.   


\section{The TESS \BP data collection}

\begin{figure*}[!h]
    \centering
    \includegraphics[width=17cm,height=4cm]{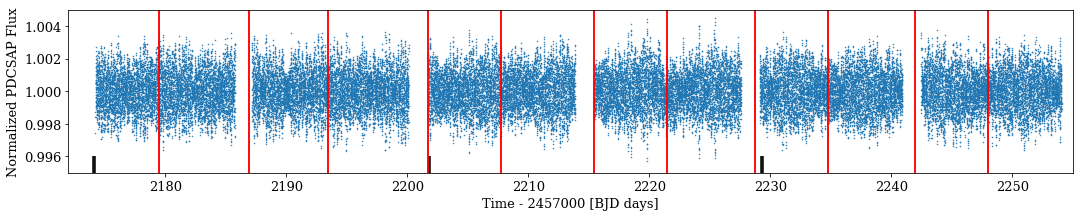}
    \caption{The stitched light curves of \BP from sectors 32, 33, and 34. The beginning of each sector is marked with a black vertical line.The red lines mark momentum dumps when angular momentum is removed from the spacecraft reaction wheels, and the spacecraft stability decreases for 10 -- 15\,min.}
    \label{fig:alldata}
\end{figure*}

The Mikulski Archive for Space Telescopes (MAST) contains observations of \BP collected by the TESS from 19 October 2018 to 8 February 2020. The TESS observed \BP in seven sectors, i.e., 4, 5, 6, 7, 32, 33, and 34, covering a period of 844 days. To analyze the \BP brightness variations, we used the 2-min ``short'' cadence Presearch Data Conditioning (PDC) light curves produced by the science analysis pipeline of the Science Processing Operation Center (SPOC; \cite{jenkins2017}). The PDC segment of the SPOC performs a set of corrections to the light curves removing the instrumental signatures, isolated outliers, correcting fluxes for the aperture effects such as field crowding or fractional loss of the target flux due to star centroid drifting \citep{jenkins2016}. In general, we used the algorithm for the light curve analysis based on the Python package ``{\sc lightkurve v2.0}'' \citep{Baren2019,lightkurve2018}, which is partially similar to that thoroughly described in \cite{zieba2019}. To assess the available data, we downloaded the fits files from the MAST archive containing the PDC fluxes, centroid measurements, and the ``quality flags'' information. The latter was used to check the possible influence of different anomalies on the flux measurements \footnote{\tiny{http://archive.stsci.edu/missions/tess/doc/TESS$_{-}$Instrument$_{-}$Handbook$_{-}$v0.1.pdf}}. Fig.~\ref{fig:alldata} shows the stitched light curves for the data set stored in sectors 32, 33, and 34. The black vertical lines show the beginning of each sector. The gaps in the light curves are caused by the science data downloading process carried out at each perigee during the 18-24 hours when no science data is collected. The red lines mark momentum dumps when the angular momentum is removed from the spacecraft reaction wheels, and the spacecraft stability decreases for 10-15 mins \footnote{\tiny{https://archive.stsci.edu/tess/tess$_{-}$drn.html}}. The data in the vicinity of the momentum dumps were omitted from the analysis.

\section{Data analysis and results}
\label{_Data}
\subsection{Identification and removal of pulsations from the \BP light curve}

\begin{figure}[!h]
    \centering
    \includegraphics[width=8cm]{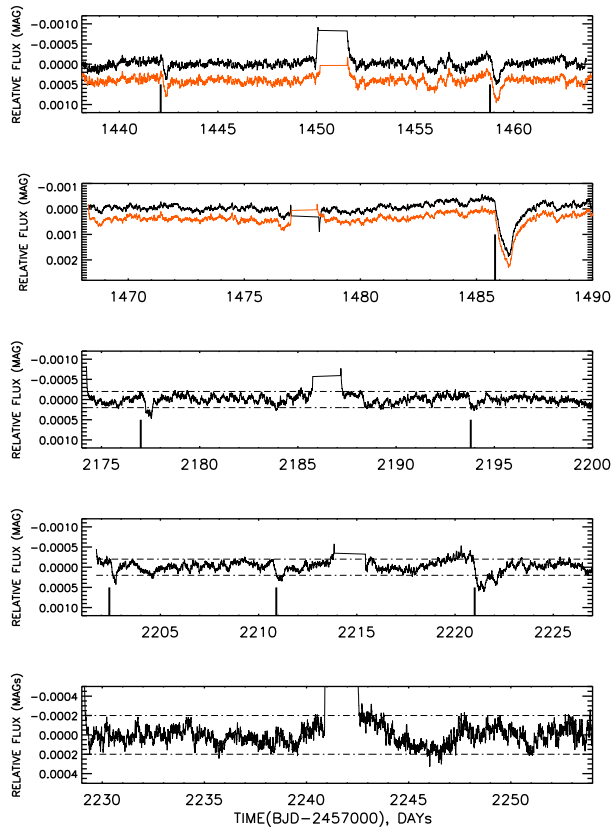}
    \caption{Differences between the model and observed light curves of \BP for sectors 5, 6, 32, 33, and 34 (from the top to the bottom, respectively). The black and orange curves in the two top panels display the light curve residuals obtained in this work and by \cite{zieba2019}, respectively. A small offset is applied for clarity.}
    \label{fig:differences}
\end{figure}

\BP belongs to a group of $\delta$ Scuti variable stars with significant periodicities of its light curve, see \cite{Koen2003,Zwintz2019}, and references therein. Therefore we remove the harmonic content from the input signal prior searching for the comet-like signatures. \cite{zieba2019} identified up to 54 significant p-modes between 23 and 76 cycles per day applying iterative fitting of the TESS \BP light curves by superposition of sine-waves extracted with the {\sc PERIOD04} program \citep{Lenz2005}. It is worth noting that the identification and removing the phase-shifting harmonics from the light curves is one of the tasks of the Transit Planet Search segment (TPS) of the science analysis pipeline, and the  whitened light curves are provided for each 2-min cadence light curve and stored in the MAST archive. The sophisticated whitening algorithms considerably simplify the transit search; however, they distort the transit shape \citep{Thompson2016}. Therefore, to avoid any degrading of the potential asymmetric dips in the light curves due to overfitting, we conducted the extraction of the harmonic signal from the light curves for each sector separately using the Python package {\sc SMURFS}, which is designed to identify and remove significant frequencies from a time series in a fully automated way\footnote{\tiny{https://github.com/MarcoMuellnher/ SMURFS}}. Thus we modeled the light curves for each sector by superposition of the pulsation frequencies between 20 and 80 cycles per day and with the amplitudes down to 0.02 mmag. For sector 6, e.g., the frequencies with S/N ratio $\ge$ 4.5 were confidently identified and consistent with those extracted by \cite{zieba2019}, see Table B.1.   Fig.~\ref{fig:differences} presents the differences between the observed and modeled light curves for  sectors 5, 6, 32, 33, and 34 observed by the TESS. For sectors 5 and 6, we also depict the residual fluxes taken from \cite{zieba2019} in order to show that applying the two different software packages, i.e., {\sc PERIOD04} and {\sc SMURFS}, to model light curves results in identical residual signals. The black lines mark the minima found in the light curves, likely associated with the transits. There are no notable dips in sector 34 which the dimming events could cause. In the bottom panel of Fig.~\ref{fig:differences} we depict the flux differences between the modeled and observed light curves of sector 34 to show that they fall in the interval $\pm$ 0.2\,mmag.


\subsection{Confirmation of the known comet transits.}

\begin{figure}[!h]
    \centering
    \includegraphics[width=7cm]{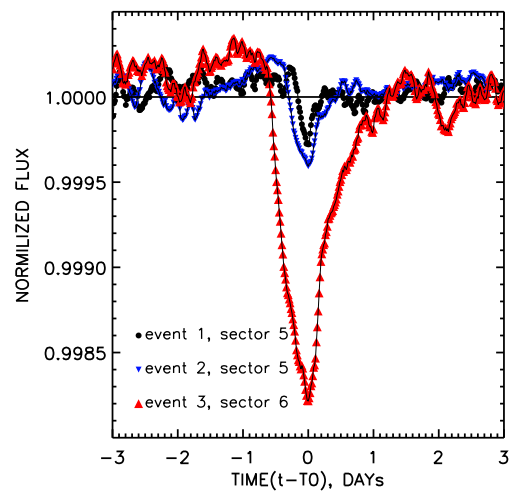}
    \caption{The asymmetric dips previously found in the TESS PDC light curves of \BP in sectors 5 and 6 \citep{zieba2019}. Time is shown relative to the moments of flux minima $T_0$ = $1442.37 \pm 0.02$ BJD, $1459.16 \pm 0.02$ BJD, $1486.40 \pm 0.01$ BJD for events 1, 2, and 3, respectively.}
    \label{fig:oldtransits}
\end{figure}

After removing the harmonic content from the signals, we averaged the flux within 30 min bins for each sector separately and smoothed the binned light curves with the Savitzky-Golay filter. The smoothing minimizes noise due to remnants of the star variability while maintaining the depth of the transits. Fig.\,\ref{fig:oldtransits} depicts the previously found minima by \cite{zieba2019} in the light curves observed in sectors 5 and 6 in the same plot. The deepest event closely resembles the theoretical prediction of the transit caused by the passage of a comet-like body across a stellar disk \citep{desEtangs1999}. The transit duration (from the beginning of the steep brightness fall to the interception with the 1.0 flux line at the transit egress) and its depth manually measured are 2.01\,$\pm$\,0.03 d and 0.99814\,$\pm$\,0.00005, respectively, which is in agreement with results presented by \cite{zieba2019}. The transit profiles depicted in Fig.\ref{fig:oldtransits} show that the normalized fluxes at the transit ingresses surpass the fluxes at the egresses for all three events. It is consistent with \cite{zieba2019}'s conclusion on the presence of the ``long-lived forward scattering halo'' associated with the deepest event, and we argue that it can also be noted in the profiles of the two other small dips.

\subsection{New comet transits}

\begin{figure}[!h]
    \centering
    \includegraphics[width=8cm]{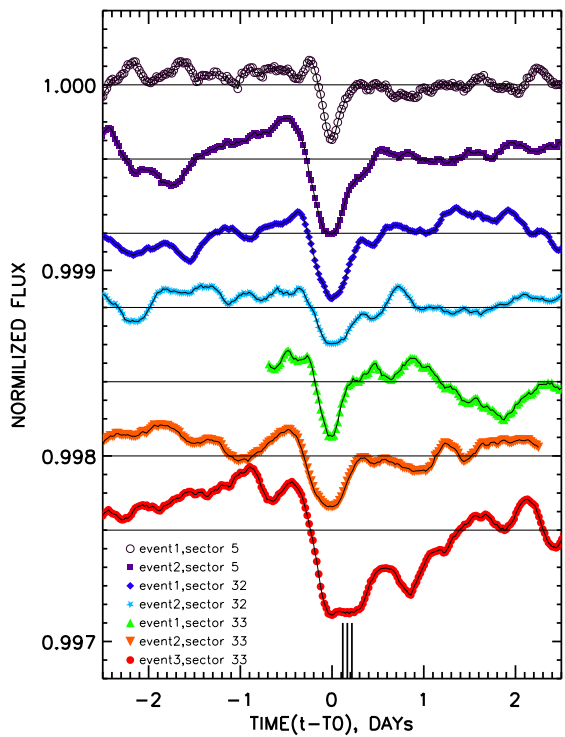}
    \caption{Asymmetric dips previously discovered in the TESS PDC light curves of BP in sector 5 \citep{zieba2019}, as well as new ones discovered in sectors 32 and 33. An offset of 0.0004 is applied on the flux values for clarity. Time is shown relative to the moments of the flux minima $T_0$: for events 1 and 2 in sector 5 is the same as in Fig.\,\ref{fig:oldtransits}; $T_0$ = $2177.45 \pm 0.03$ BJD and $2193.98 \pm 0.02$ BJD for events 1 and 2 in sector 32, and $T_0$ = $2202.71 \pm 0.02$ BJD, $2211.11 \pm 0.01$ and $2221.4 \pm 0.1$ BJD for events 1, 2, and 3 in sector 33, respectively.}
    \label{fig:newtransits1}
\end{figure}

\begin{figure*}
    \centering
    \includegraphics[width=17cm]{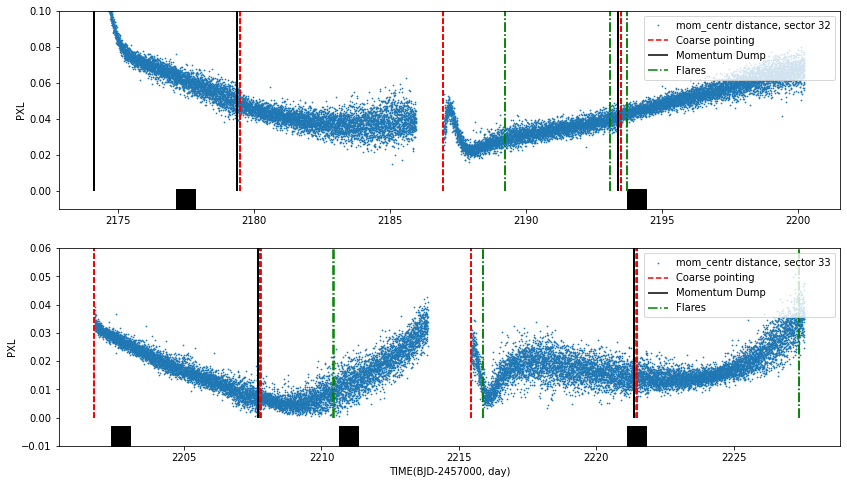}
    \caption{The instantaneous positions of the target's flux-weighted centroids for sectors 32 and 33. The black rectangles mark the transit events identified. The red vertical dashed lines and black solid lines mark the intervals of the spacecraft's coarse pointing and the momentum dump events, respectively. The green dotted-dashed lines mark flares.}
    \label{fig:mom_dumps33_last}
\end{figure*}

Five new asymmetric non-periodic minima in the \BP light curve were identified in sectors 32 and 33. The shapes of the new dips are similar to those shallow features observed in sector 5. Meanwhile, the light curve downloaded from sector 34 is featureless. Fig.\,\ref{fig:newtransits1} shows the profiles of new events in sectors 32 and 33, along with the previously observed ones in sector 5 for comparison. The first dip detected in sector 33 occurred at the beginning of the sector (the minimum corresponds to BJD,  Barycentric Julian Date\, $T_0$ = $2202.71\pm 0.02$) so this light curve in the figure is truncated from the left. We estimated the duration and depth of all dips with manual measurements of the time span between the ingress and egress points on the smoothed light curves presented in  Fig.\,\ref{fig:oldtransits}, Fig.\,\ref{fig:newtransits1}. The measured duration values are listed in Table\,\ref{table:model_parameters} (see section 3.3). All weak dips are very shallow at the level of 0.03\,\% with a duration between 0.5 and 0.9\,days. The only exception is the last event identified in sector 33, which is not reliable because it overlaps partially with the spacecraft's coarse pointing due to the momentum dump: the transit ingress starts at approximately 2220.74 BJD, the flux minimum occurs at approximately 2221.4 BJD, and the momentum dump event covers the time span between 2221.46 and 2221.47 BJD. To verify the reliability of the data from sectors 32 and 33, we examined the instantaneous row and column positions of the target's flux-weighted centroids in the CCD frames. This information is available for each 2-min cadence PDCSAP light curve. Additionally, we looked through TESS Data Release Notes on sectors 32 and 33, which contain information about observation circumstances, notes on individual targets, spacecraft pointing, data anomalies, etc. \footnote{\tiny{http://tasok.dk/info/docs.php}}. Fig.\,\ref{fig:mom_dumps33_last} shows the position drift of the flux-weighted centroids with time. The black rectangles mark the identified transit events, the red dashed lines mark coarse spacecraft pointing, mainly due to the momentum dumps (solid black lines), and the green dash-dotted lines show flares in the frames, which can degrade the background signal. Although the last event overlaps the spacecraft's coarse pointing period due to the momentum dump, the centers of the star centroid did not shift significantly on the CCD detector ($\le$ 0.01 pixel, negligible compared to the aperture size) to cause the fractional loss of the target flux.

\subsection{Modeling of transit events}

To model exocomet transits we followed the procedure described in \cite{brogi2012} and \cite{zieba2019}. In fact, in our computation we used the modification of the Python code uploaded to GitHub by \cite{zieba2019}\footnote{\tiny{https://github.com/sebastian-zieba/betaPic$\_$comet}}.
Bellow we provide some information about the main steps and essential features of the model calculations.

-- We assume a circular comet orbit with period {\it P} orbiting the star of mass {\it M} and radius {\it R}. For \BP we adopted $M_\mathrm{star} = 1.80M_\mathrm{Sun}$ and $R_\mathrm{star} = 1.53R_\mathrm{Sun}$ \citep{Wang2016}.

-- Light curves with removed oscillations by the {\sc SMURFS} package were used in our analysis. We compared our cleaned light curves with the light curves reduced by \cite{zieba2019} and found rather marginal differences, see Section 3.1.

-- \cite{zieba2019} used five parameters ($t_\mathrm{mid}$ , {\it b}, $c_\mathrm{e}$, $\lambda$, {\it P}) to optimize the model. Here $t_\mathrm{mid}$ marks an arbitrary point along the orbit  following the nomenclature of \cite{brogi2012}, but we adopt it as the midpoint of the transit, {\it b} is the impact parameter, $c_\mathrm{e}$ and $1/\lambda$ determine the maximum of extinction cross section and characteristic size scale of the cometary tail, $P$ is the orbital period (see Appendix\,\ref{appendix:formulae} for more details).

-- We used the Python routine {\sc scipy.optimize} to determine a zero approximation for the input parameter set by the
chi square minimisation to fit the observed transit profile. The parameter estimates  serve as the input set for a Markov-Chain Monte Carlo ({\sc MCMC}) fit (see below). Following \cite{zieba2019} we fixed $b = 0$ to avoid degeneration of the solution caused by the free {\it b} parameter. In that way, we constrain the range of {\it P} and $\lambda$ selection.

-- As noted above, the deepest transit in sector 6 and some of the shallow transits show a higher flux level at their ingress points compared to the egresses. Therefore, we carried out the fitting procedure for two cases, i.e., with and without a scattering function in the model. To account for dust scattering, we used the approach proposed by \cite{brogi2012}, but 
we did not treat the parameter {\it g} ($g \simeq \cos(\varphi)$, where $\varphi$ is the cone angle that beams scattering) as a free one, but rather adjusted it iteratively while obtaining the zero approximation set of input parameters.
After that, we kept {\it g} parameter fixed not including it into the input set for the {\sc MCMC} fit. To account for scattering, we extent the region of the light curve to which we fit the model up to approximately $\pm$\,3 days around the $t_\mathrm{mid}$.  

-- We used the Python software package {\sc emcee} \citep{Foreman-Mackey2013} to perform a {\sc MCMC} fit to the binned light curve around the transit features.

The best fit parameters of our model calculations are listed in Table\,\ref{table:model_parameters} and the transit profiles modeled with the set of the best-fit parameters are also depicted (see  Appendix\,\ref{appendix:figures}). We used the orbital periods derived to calculate the orbital axes assuming circular orbits, which are also presented in Table\,\ref{table:model_parameters} along with the manually measured durations of the transit events.

\begin{table*}
\caption{Fitting parameters from {\sc MCMC}. Indices at the sector number in the first column specify the event number. $T_\mathrm{dur}$ is an event duration measured manually.}
\label{table:model_parameters}      
\centering          
\begin{tabular}{cccccccccc}
\hline\hline       
\multirow{2}{*}{Sector} & \multirow{2}{*}{$T_\mathrm{0}$} & \multirow{2}{*}{{\it g}} & \multirow{2}{*}{$C_\mathrm{e}\times10^{3}$} & 1000/$\lambda$, & \multirow{2}{*}{{\it P}, days}  & $V_\mathrm{circ}$, & $a_\mathrm{circ}$, & $T_\mathrm{dur}$,\\
                        &                                 &                          &                                             & 1/rad           &                                 & km/s               & au                 & days\\  
\hline                    
    5$_{1}$                     & 1442.357$\pm$0.004 & 0.9975 & 1.17$\pm$0.14 & 21.7$\pm$7.0  & 625  & 30.2$\pm$1.0   & 1.74 & \multirow{2}{*}{0.60$\pm$0.05} \\  
                                & 1442.425$\pm$0.010 & 0      & 1.15$\pm$0.80 & 2.4$\pm$4.0   & 151  & 48.6$\pm$11.1  & 0.68 & \\  
    5$_{2}$                     & 1458.970$\pm$0.003 & 0.9970 & 1.58$\pm$0.07 & 6.6$\pm$0.6   & 811  & 27.8$\pm$0.4   & 2.07 & \multirow{2}{*}{1.04$\pm$0.05}\\  
                                & 1459.020$\pm$0.003 & 0      & 1.16$\pm$0.11 & 2.5$\pm$0.5   & 360  & 36.3$\pm$1.6   & 1.21 &\\  
    6\tablefootmark{*}          & 1485.919$\pm$0.001 & 0.9950 & 3.73$\pm$0.03 & 1.23$\pm$0.03 & 2753 & 18.48$\pm$0.08 & 4.68 & \multirow{2}{*}{2.01$\pm$0.03}\\  
                                & 1485.947$\pm$0.002 & 0      & 3.83$\pm$0.03 & 0.99$\pm$0.03 & 2335 & 19.52$\pm$0.14 & 4.19 & \\ 
    6\tablefootmark{**}         & 1486.290$\pm$0.001 & 0      & 3.67$\pm$0.04 & 1.22$\pm$0.04 & 2307 & 19.6$\pm$0.1   & 4.15 & --\\  
    32$_{1}$                    & 2177.397$\pm$0.005 & 0.9950 & 1.18$\pm$0.05 & 3.0$\pm$0.6   & 640  & 30.1$\pm$0.9   & 1.76 & \multirow{2}{*}{0.77$\pm$0.11} \\  
                                & 2177.498$\pm$0.005 & 0      & 2.22$\pm$0.05 & 0.7$\pm$0.2   & 481  & 33.0$\pm$0.8   & 1.46 & \\  
    32$_{2}$                    & 2193.890$\pm$0.003 & 0.9975 & 1.18$\pm$0.11 & 5.5$\pm$1.0   & 735  & 28.7$\pm$0.8   & 1.94 & \multirow{2}{*}{0.85$\pm$0.05} \\  
                                & 2193.920$\pm$0.005 & 0      & 0.94$\pm$0.30 & 1.2$\pm$0.9   & 441  & 34.0$\pm$3.1   & 1.38 &\\  
    33$_{1}$                    & 2202.646$\pm$0.003 & 0.9700 & 0.76$\pm$0.03 & 15.1$\pm$1.8  & 124  & 52.0$\pm$1.5   & 0.59 & \multirow{2}{*}{0.51$\pm$0.04} \\ 
                                & 2202.673$\pm$0.003 & 0      & 0.99$\pm$0.10 & 4.8$\pm$1.1   & 106  & 54.8$\pm$2.8   & 0.53 &\\  
    33$_{2}$                    & 2210.931$\pm$0.003 & 0.9800 & 0.50$\pm$0.01 & 5.5$\pm$0.5   & 743  & 28.6$\pm$0.6   & 1.95 & \multirow{2}{*}{0.67$\pm$0.02}\\  
                                & 2210.947$\pm$0.006 & 0      & 0.47$\pm$0.03 & 9.5$\pm$5.8   & 252  & 41.0$\pm$5.5   & 0.95 &\\  
    33$_{3}$\tablefootmark{***} & 2221.416$\pm$0.004 & 0.9955 & 1.19$\pm$0.02 & 5.0$\pm$0.3   & 1802 & 21.3$\pm$0.2   & 3.53 & \multirow{2}{*}{1.76$\pm$0.13}\\  
                                & 2221.510$\pm$0.003 & 0      & 0.80$\pm$0.02 & 3.2$\pm$0.4   & 918  & 26.6$\pm$0.2   & 2.25 &\\  
\hline                  
\end{tabular}
\tablefoot{The first line of each transit contains model parameters if a scattering function is included in the model, and the second line contains model parameters if scattering is not taken into account.\\
\tablefoottext{*}{this work} \\
\tablefoottext{**}{\cite{zieba2019}}\\
\tablefoottext{***}{degraded by the  momentum dump}
}
\end{table*}


\section{Discussion}

The asymmetric shallow dips found in sectors 32 and 33 have a depth and duration similar to those found in sector 5. These drops in the star flux could not be explained by cold spots on the star's surface, as they do not show periodicity connected to the \BP rotation period ($\sim$16 hours, $v \sin(i) = 130$\,km\,s$^{-1}$, $i\sim90^{\circ}$, see \cite{roye07}) and could not disappear in 1--2 days. It has already been noted that the brief variations of the \BP flux were discussed in the context of either a dusty cloud passing over the star disk or an exocomet transit; both explanations seem to be plausible \citep{Lamers1997}. In both cases, the authors consider the contribution of scattering to take into account the brightness spikes before and after the drop of the star flux. In the case of a dusty cloud, the phase function of dust particles can be strongly peaked forward due to diffraction, with the parameter {\it g} reaching 0.98 if the cloud is located at a distance larger than 1.5\,AU \cite{Lamers1997}. Assuming circular orbits, the orbital periods obtained with the model allow us to estimate the distances from the star. If the scattering function is not included in the model, the orbital distances of the transiting objects are between 0.5 and 1.5\,AU (we do not discuss the last event in sector 33, for which the star flux was degraded by the spacecraft's coarse pointing). Including the scattering function into the model influences mostly the parameter describing the characteristic size scale of the cometary tail, $\lambda$, increasing the orbital period and orbital semimajor axis values. Because the scattering function is strongly peaked around the transit midpoint, the model parameters change depending on the it g value and transit profile characteristics such as asymmetry relative to the transit midpoint and depth at the flux minimum. According to the model parameters listed in Table \ref{table:model_parameters}, including the scattering function in the model has a significant impact on the calculated orbital period and circular velocity for events 1 in sector 5 and 2 in sector 33; both of which have slightly shallower profiles than the others.

The parameters extracted for the deepest transit (in sector 6) weakly depend on the scattering function, and in the case of $g=0$ the model results are in reasonable agreement with those presented in \cite{zieba2019}. For all events considered, the increase of the total flux (the star + comet) at the ingress points is approximated by the scattering function with a very high {\it g} parameter, pointing out the very narrow scattering that can take place if the medium consists of a mixture of diffracting dusty particles of different sizes \cite{Lamers1997}.

It is also worth noting that several groups can be identified among the \BP FEB events based on the peculiarities of the spectral features observed \citep{Beust2000}. These groups consist of bodies that evaporate, likely at different distances from the star. Modeling of the TESS light curves and the transit duration intervals shows that the deepest and longest events in sector 6, and possibly, the last one in sector 33, with a duration of about 2\,days are caused by bodies orbiting the star at distances larger than at least 2\,AU. Meanwhile, the comet-like objects that move closer to the star are responsible for the shallow events.  

\cite{Rap2018} report the first strong evidence for exocomet transits based on the analysis of the light curves of two F2V stars, KIC\,3542116 and KIC\,11084727 from the $Kepler$ database. Among seven events discovered in the star's light curve, three are very shallow ($\leq$ 1\%) with a duration of less than 1 day. Another star with an exocomet signature is KIC 8462852, whose light curve observed with the Kepler mission reveals irregularly shaped aperiodic dips of different depths \citep{Boyaj2016}. The plausible models to explain the observed light curves invoke circumstellar material spread around a single elliptical orbit or the fragmentation of one massive exocomet into multiple cometary nuclei with slightly different orbits \citep{Boyaj2016,Bodman2016,wyat18}. The sporadic drops in the star's light curve can also be linked to the phenomenon of "little dippers," which are quasi-periodic or aperiodic minima in the observed light curves of young stars (less than 10 Myr) or even comparatively old stars without IR excess, and were first discovered in the classical T-Tauri star DF Tau \citep{ chelli1999}.The drops in the light curves have diverse shapes and durations of $\sim$0.5\,-- 1\,days and decrease of $\sim$0.1\,-- 1.0\% in the flux (see \cite{bodm17,ansd19}, and references therein). Some quasi-periodic variations in the ``little dipper'' light curves can be explained by a model of the spots on the surface of rotating stars or by extinction caused by optically thick dusty clamps co-rotating with the star \citep{bodm17}. But other irregular dimming events are also consistent with the transit of a comet-like body in a circular or eccentric orbit \citep{scar2016}.

In this paper, we performed the analysis of light curves of \BP observed by TESS in seven sectors and detected 8 events of our interest, see Table \ref{table:model_parameters}. Thus, we have an average of 1 event every 27 days, but their distribution is very uneven. Namely, in the sectors 4, 7, and 34, such events are not observed at all; on the other hand, in sector 33 we found as many as 3. Thus far, we can claim that weak dimming events will be observed more often. It is worth noting that the irregularity of manifestations of related objects (FEBs) has been claimed in some works (see, for example, Beust2000), but the frequency of events is much higher, reaching several hundred per year.


\section{Conclusions}

We independently analyzed the light curves of \BP observed by the TESS mission to perform a new search for exocomet transits in the recently observed sectors 32, 33, and 34. We do not detect any regular planet transits in the 2 minute short cadence light curves. We confirm the existence of the known comet transits detected by \cite{zieba2019}. We report new transit events in sectors 32 and 33 that have asymmetric profiles with very shallow flux drops at the level of 0.03\,\% and duration between 0.5 and 2.00\,days with a measurement uncertainty of 0.03 -- 0.13\,days.

The newly found transits are very similar in duration and depth to those shallow asymmetric dips (except the last one that is degraded by the spacecraft coarse pointing), which were discovered by \cite{zieba2019} in sector 5, as well as to those found for KIC 3542116 and KIC 11084727 in the Kepler database \citep{Rap2018}.

We conclude that the young system of \BP shows significant comet activity. At least 8 events with asymmetrical profiles resembling the shapes of exocomet transits have been detected in a comparatively short time interval of $\sim$850\,days, among which at least one deep transit has been detected. In the old Solar system, a big comet appeared once in 10\,-- 20\,years, see \cite{lich99}.

In the case of \BP, we provide more evidence for the similarity of the newly discovered transits with the known ones with a depth of less than 0.1\,\%; most likely, it is a common phenomenon.

\begin{acknowledgements}
      This study was performed in the frames of the government funding program for institutions of the National Academy of Sciences of Ukraine (NASU) and supported by the National Research Foundation of Ukraine (\textnumero\,2020.02/0228). The authors gratefully thank to the anomymous Referee for the constructive comments and recommendations which definitely help to improve the readability and quality of the paper. We thank Dr. Sebastian Zieba for his comments and help with the model calculations. All data presented in this paper were obtained from the Mikulski Archive for Space Telescopes (MAST), which is hosted at STScI. STScI is operated by the Association of Universities for Research in Astronomy, Inc., under NASA contract NAS5-26555. Support for MAST for non-HST data is provided by the NASA Office of Space Science via grant NNX09AF08G and by other grants and contracts. This paper includes data collected by the TESS mission. Funding for the TESS mission is provided by the NASA Science Mission directorate.
      Many thanks to Prof. Hugh Jones (UH, Hatfield) for his help in improving the language of the paper.
\end{acknowledgements}

\bibliographystyle{aa} 
\bibliography{Pavlenko} 
%

\begin{appendix} 
\section{Comet transit modeling formulae.}
\label{appendix:formulae}

 To calculate the theoretical profiles we use the model described 
by \cite{brogi2012} and adapted by \cite{zieba2019}. 
It is considered that a comet moves on a circular orbit,  
see Fig.\,4 in \citet{brogi2012} for more geometric details. 

The total flux as a function of orbital phase $\varphi$ is 
$$
  I(\varphi) = \frac{1}{\delta}\int_{2\pi\varphi-\delta/2}^{2\pi\varphi+\delta/2} [ 1 - I_\mathrm{e}(\theta) + I_\mathrm{s}(\theta) ]d\theta,
$$
\noindent where $I_\mathrm{e}$ and $I_\mathrm{s}$ are the contributions of the extinction and scattering components, respectively. $\varphi=2\pi\theta$, here $\theta$ is the angle between the observer, the centre of the star and the orbiting body. $\delta = 2\pi\Delta t/P$, where $\Delta t$ is the exposure time, and $P$ is the comet's orbital period.

 The extinction component is expressed as
$$
  I_\mathrm{e}(\theta) = \int_{0}^{2\pi}\rho(\theta - \theta^{\prime})i(\theta^{\prime},\hat{r}_\mathrm{c})d\theta^{\prime},
$$
\noindent where $\theta - \theta^{\prime} = \Delta\theta$ is the angular distance between the position of the comet and the arbitrary point along the orbit (in this work, $\theta^{\prime}$ marks the midpoint of the transit event).

The model assumes an optically thin cometary coma expressed in units of the stellar area and its extinction cross-section $\rho$ drops exponentially away  from the center as
$$
  \rho(\Delta\theta) = \frac{\rho_\mathrm{0}}{\pi R^{2}_{\star}} e^{-\lambda(\Delta\theta)} \equiv c_\mathrm{e}e^{-\lambda(\Delta\theta)},
$$
\noindent where the multiplicative factor, $c_{e}$, and the exponential parameter, $\lambda$, are free parameters of the model. 

The chord length that a comet traverses the stellar disk is represented by the impact parameter $b = [1-(r_\mathrm{c}/2R_\mathrm{\star})^{2}]^{1/2}$. The angle of the crossed chord as seen from the comet orbit is given by $\hat{r}_\mathrm{c} = \arcsin(r_\mathrm{c}/2a)$. Here $R_\mathrm{\star}$ and {\it a} are the stellar radius and the orbital radius of the comet, respectively. The latter is calculated from the orbital period.

The extinction component is the convolution between the extinction cross-section and the intensity of the stellar disk. The latter one, in terms of $\theta^{\prime}$ and $\hat{r}_\mathrm{c}$, is given by
$$
  i(\theta^{\prime},\hat{r}_\mathrm{c}) = 1 - u\left[ 1 - \frac{a}{R_\mathrm{\star}}\sqrt{\sin^{2}(\hat{r}_\mathrm{c}/2) - \sin^{2}\theta^{\prime}} \right],
$$
\noindent where {\it u} is the linear limb-darkening coefficient, which we adopt as 0.79 \citep{claret2000}.

The scattering component $I_\mathrm{s}(\theta)$ is given by
$$
  I_{s}(\theta) = \pi\varpi\left( \frac{R_\mathrm{\star}}{a} \right)^{2}\int^{2\pi}_{0} \rho(\theta - \theta^{\prime})\bar{p}(\theta^{\prime}) d\theta^{\prime},
$$
\noindent where $\bar{p}(\theta^{\prime})$ is the Henyey-Greenstein (H-G) phase function \citep{Henyey1941}
$$
  \bar{p}(\theta^{\prime}) = \frac{1 - g^{2}}{4\pi(1 - 2g\cos\theta^{\prime} + g^{2})^{\frac{3}{2}}}.
$$
\noindent The parameter {\it g} is the asymmetry parameter in the range between -1 and 1 with $g > 0$ and $g < 0$ for forward-peaked and backward-peaked scattering functions, respectively, and $\varpi$ is the single-scattering albedo. The latter parameter we fix at 0.15, taking into account the estimation of single-scattering albedo values for the solar system comets and assuming a cometary coma as a mixture of volatile and solid dusty components \citep{meech1987,frattin2017}.   

\section{Modeled exocometary transits in \BP system} 
\label{appendix:figures}

This appendix contains the binned \BP light curves limited to approximately $\pm$ 3 days around the transit events, which were used for the model calculation of the transit profiles with {\sc MCMC}. For each transit, the model profile calculated with the set of the best-fit parameters is also depicted. The left panel of each figure presents the modeled light curve if the scattered function is not included in the model, whereas the right panel shows the transit profile if scattering is taken into account. The sector in which the event occurs as well as the event number are specified in the figure caption. The bottom panels depict residuals.

\begin{figure}[!h]
    \centering
    \includegraphics[width=0.48\linewidth]{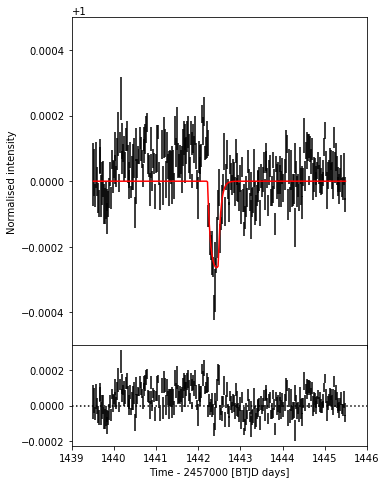}
    \includegraphics[width=0.48\linewidth]{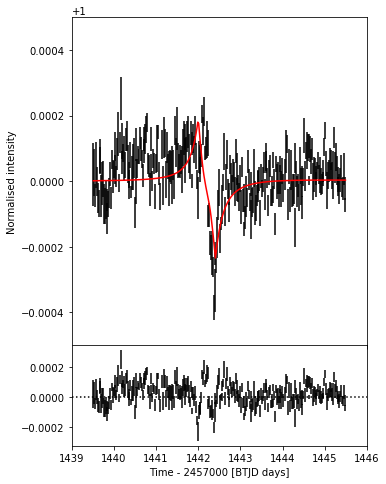}
    \caption{Sector 5, event 1.}
    \label{fig:transit_model_5a}
\end{figure}
\begin{figure}[!h]
    \centering
    \includegraphics[width=0.48\linewidth]{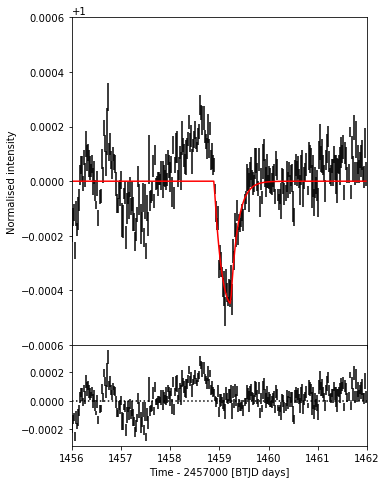}
    \includegraphics[width=0.48\linewidth]{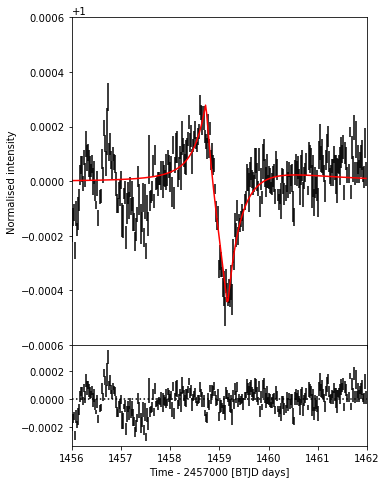}
    \caption{Sector 5, event 2.}
\label{fig:transit_model_5b}
\end{figure}
\begin{figure}[!h]
    \centering
    \includegraphics[width=0.48\linewidth]{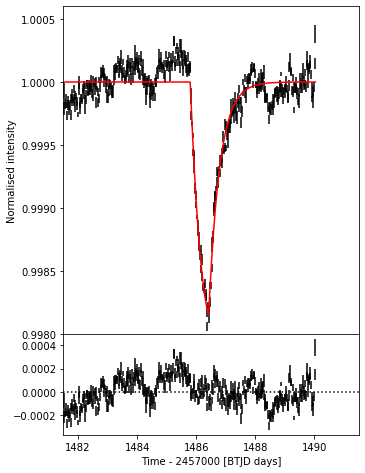}
    \includegraphics[width=0.48\linewidth]{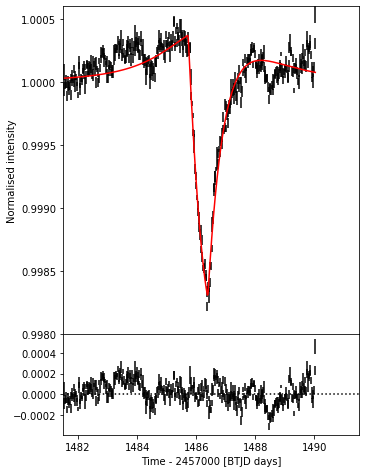}
    \caption{The transit in sector 6.}
    \label{fig:transit_model_6}
\end{figure}
\begin{figure}[!h]
    \centering
    \includegraphics[width=0.48\linewidth]{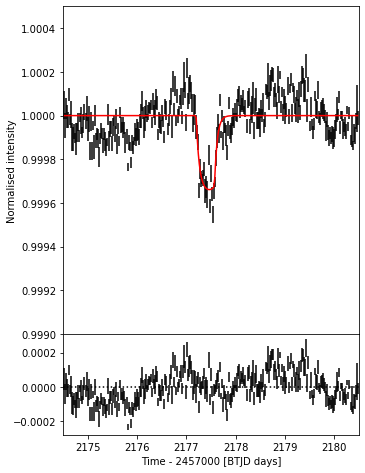}
    \includegraphics[width=0.48\linewidth]{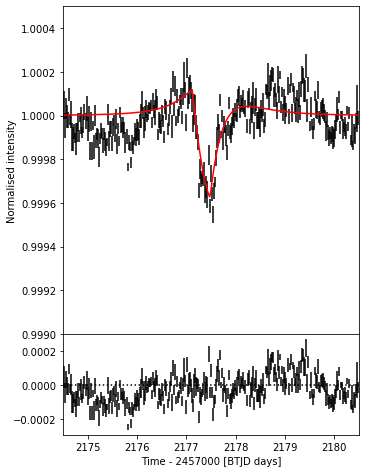}
    \caption{Sector 32, event 1.}
    \label{fig:transit_model_32a}
\end{figure}
\begin{figure}[!h]
    \centering
    \includegraphics[width=0.48\linewidth]{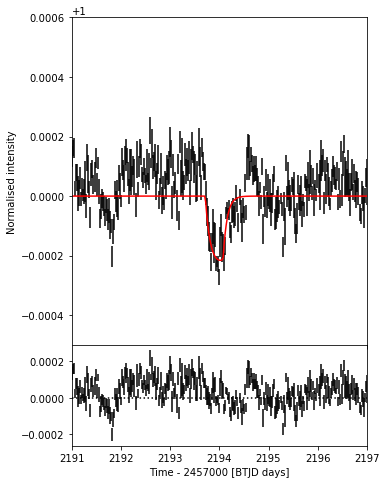}
    \includegraphics[width=0.48\linewidth]{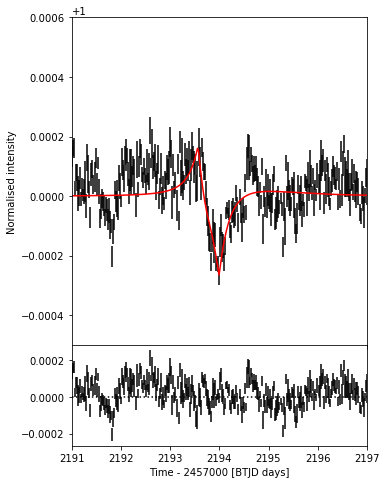}
    \caption{Sector 32, event 2.}
    \label{fig:transit_model_32b}
\end{figure}

\begin{figure}[!h]
    \centering
    \includegraphics[width=0.48\linewidth]{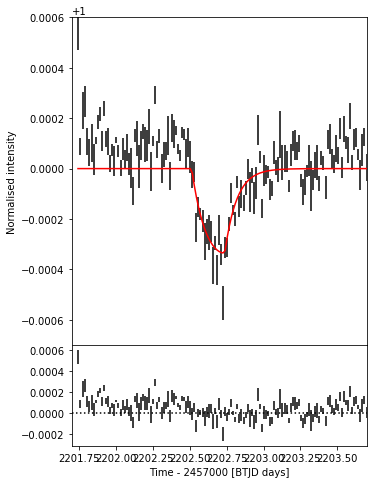}
    \includegraphics[width=0.48\linewidth]{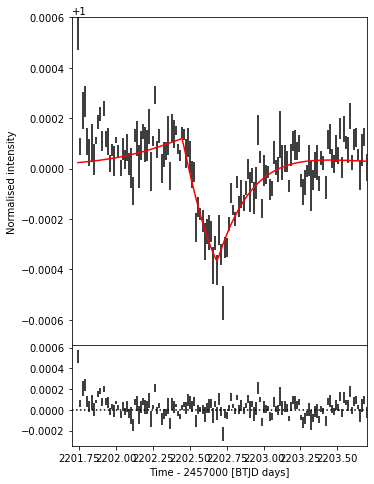}
    \caption{Sector 33, event 1.}
    \label{fig:transit_model_33a}
\end{figure}
\begin{figure}[!h]
    \centering
    \includegraphics[width=0.48\linewidth]{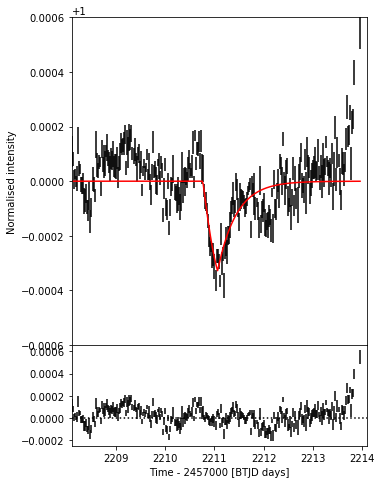}
    \includegraphics[width=0.48\linewidth]{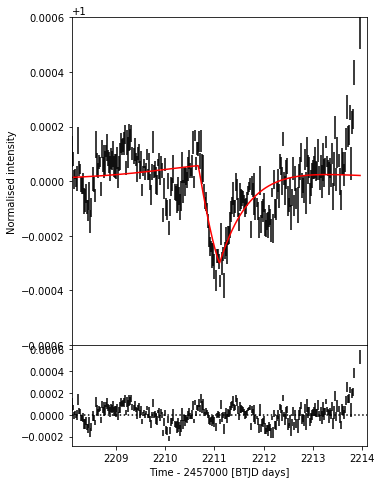}
    \caption{Sector 33, event 2.}
    \label{fig:transit_model_33b}
\end{figure}
\begin{figure}[!h]
    \centering
    \includegraphics[width=0.48\linewidth]{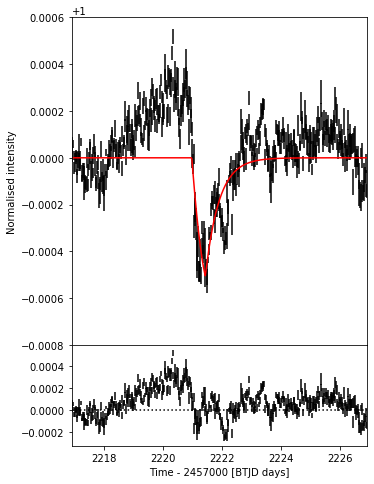}
    \includegraphics[width=0.48\linewidth]{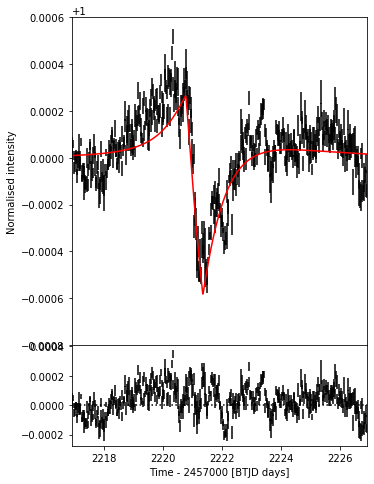}
    \caption{Sector 33, event 3.}
    \label{fig:transit_model_33c}
\end{figure}

\end{appendix}

\end{document}